\documentclass[twocolumn,superscriptaddress,showpacs,preprintnumbers,amsmath,amssymb]{revtex4}

\newcommand{\bb}{\begin{equation}}
\newcommand{\en}{\end{equation}}

\usepackage{graphicx}

\begin{document}

\title{Fluctuation Theorem for the flashing ratchet model of molecular motors}
\author{D. Lacoste}
\affiliation{Laboratoire de Physico-Chimie Th\'eorique, UMR 7083,
ESPCI, 10 rue Vauquelin, 75231 Paris cedex 05, France}
\author{K. Mallick}
\affiliation{Institut de Physique Th\'eorique, CEA Saclay, 91191
Gif, France}

\date{\today}
\begin{abstract}
Molecular motors convert chemical energy derived from the hydrolysis of ATP
into mechanical energy. A well-studied model of
a molecular motor is the flashing ratchet model. We show that this model
exhibits a fluctuation relation known as the Gallavotti-Cohen symmetry.
Our study highlights the fact that the symmetry is present only if
the chemical and mechanical degrees of freedom are both included in the description.
\end{abstract}
 \maketitle

Molecular motors are subject to intense study both from
biological and technological point of view \cite{howard,hanggi}.
These remarkable nanomachines are enzymes capable of converting
chemical energy derived from ATP hydrolysis into mechanical work.
They typically operate far from equilibrium, in a regime
where the usual thermodynamical laws do not apply.
Generically such motors are modelled either in terms of continuous
flashing ratchets \cite{armand1,reiman} or by a master
equation on a discrete space \cite{kolomeisky_revue,lipowsky}.
Recently, a general organizing principle for non-equilibrium systems has emerged
which is known under the name of fluctuation relations \cite{kurchan,ritort-2005}.
These relations, which hold for non-equilibrium steady states, can be
seen as macroscopic consequences of generalized detailed balance conditions, which themselves arise due to
the invariance under time reversal of the dynamics
at the microscopic scale \cite{derrida-2007}.

An interesting ground to apply these concepts is the field of
molecular motors \cite{gaspard_andrieux,seifert,lipowsky_JSTAT,
prl,pre,maes,schmiedl-2007}.
The fluctuation relations impose thermodynamic constraints
on the operation of these machines, particularly in
regimes arbitrary far from
equilibrium. Near equilibrium, they lead to Einstein
and Onsager relations. For non-equilibrium steady states, they can be used
to quantify deviations from Einstein and Onsager relations as we have shown in
Refs.~\cite{prl,pre}.

In this paper, we investigate fluctuation relations for continuous ratchet models.
We first study a purely mechanical ratchet (model I),
which applies to the translocation of a polymer through a pore \cite{dlubensky}.
We then consider a flashing ratchet (model II),
which applies to molecular motors \cite{armand1}.
Using a method inspired by Refs.~\cite{kurchan,lebowitz}, we show that the Gallavotti-Cohen symmetry is always present in model I, but we emphasize that in model II the symmetry exists only if the chemical and mechanical degrees of freedom of the motor are both included in the description.

%
%
Let us first consider a random walker in a periodic potential subject to an external force $F$ (model I) \cite{risken,hanggi}. The corresponding Fokker-Planck equation is
\begin{equation}
   \frac{\partial P}{\partial t} = D_0 \frac{\partial}{\partial x} \left[
  \frac{\partial P}{\partial x}  +
 \frac{{U'}(x) - F }{k_B T} P \right],
 \label{eq:FPuncoupled}
\end{equation}
where $U(x)$ is a periodic potential $U(x + a) =  U(x)$ and $a$ is the period.
This equation describes the stochastic dynamics of a particle
 in the effective potential $U_{eff}(x)
  = U(x) - Fx$.
  By solving Eq.~(\ref{eq:FPuncoupled}) with periodic boundary conditions \cite{dlubensky,risken}, it can be readily proven that
  the system reaches a stationary state with
 a uniform current $J$ in the long time limit. This current is
 non-vanishing if a non zero force is applied. When $F=0$, there is no tilt in the potential, $J=0$ and the stationary probability is given by the equilibrium Boltzmann-Gibbs factor.

  We  call $x(t)$ the position of the ratchet at time $t$ knowing that
the ratchet was located at $x(0) = 0$ at time $t=0$, which we
decompose as $x = (n + \zeta)a$ where
$n$ is an integer and $ 0 \le \zeta  < 1$.
 The stationary current $J$ is related to the average position $x(t)$ by
$J =  \lim_{t \to \infty} \frac{\langle x(t) \rangle}{t},$
i.e. $J$ is the mean speed of the ratchet in the long time limit.
 More generally  we are  interested in
  the higher cumulants of  $x(t)$
 when $t \to \infty$.
It is useful to define the generating function
  \begin{equation}
   F_{\lambda}(\zeta, t) = \sum_n \exp\left( \lambda(\zeta +n) \right)
 P( (n+\zeta)a, t)   \, .
\label{GenFunct}
 \end{equation}
The time evolution of this
generating function $F_{\lambda}$ is obtained by
summing over Eq.~(\ref{eq:FPuncoupled}). This leads to the following equation:
 \begin{equation}
  \frac{\partial F_{\lambda}(\zeta, t)} {\partial t }
 =   {\mathcal L}(\lambda)  F_{\lambda}(\zeta, t)   \, ,
\label{EvolGenFunct}
 \end{equation}
 where the  deformed differential operator ${\mathcal L}(\lambda)$ acts
 on a periodic function $\Phi(\zeta,t)$ of period $1$
 as follows:
  \begin{equation}
 \frac{a^2}{D_0} {\mathcal L}(\lambda)  \Phi =
 \frac{\partial^2 \Phi } {\partial \zeta^2 } +
 \frac{\partial }{\partial \zeta}  \Big(
   \tilde{U}'_{eff} \Phi  \Big)
 - 2 \lambda  \frac{\partial \Phi } {\partial \zeta  }
   -  \lambda \tilde{U}'_{eff}  \Phi + \lambda^2 \Phi \, ,
\label{def:Ldeformed}
  \end{equation}
where $\tilde{U}'_{eff}=a \partial_x U_{eff}/k_B T$ and the left hand side of Eq.~(\ref{def:Ldeformed}) is proportional to the inverse of the characteristic time $\tau=a^2/D_0$. A similar procedure exists in solid state physics, where periodic functions
are expanded in eigenfunctions of Bloch form, which are
 eigenfunctions of an operator similar to
${\mathcal L}(\lambda)$ \cite{dlubensky}.

 The operator ${\mathcal  L}(\lambda)$  has the following
 fundamental  conjugation property:
\begin{equation}
  \hbox{e}^{U(x)/k_B T}
   {\mathcal  L}(\lambda)  \left( \hbox{e}^{-U(x)/k_B T}
   \Phi \right) =
  {\mathcal L^\dagger} \left(-f -\lambda \right)  \Phi,
 \label{OperatorGC1}
  \end{equation}
  with $f=Fa/k_B T$ the normalized force. This property implies that
  operators  ${\mathcal L}(\lambda)$
 and  ${\mathcal L^\dagger} \left(-f -\lambda \right)$ are
 adjoint to each other, and thus have the same spectrum.
If we call $\Theta(\lambda)$ the largest eigenvalue of
${\mathcal L}(\lambda)$,
we obtain from Eq.~\ref{OperatorGC1}
that $\Theta(\lambda)$ satisfies the Gallavotti-Cohen symmetry:
\begin{equation}
 \Theta(\lambda) =  \Theta(-f -\lambda).
\label{GC1}
 \end{equation}
In fact, this symmetry holds for all eigenvalues.
 For the special  case $f=0$, the conjugation relation~(\ref{OperatorGC1})
 reduces to the {\it detailed balance} property \cite{lebowitz}.
Finally, it is important to note that $\Theta(\lambda)$ is the generating
 function for the cumulants of $x(t)$.

\begin{figure}
\includegraphics{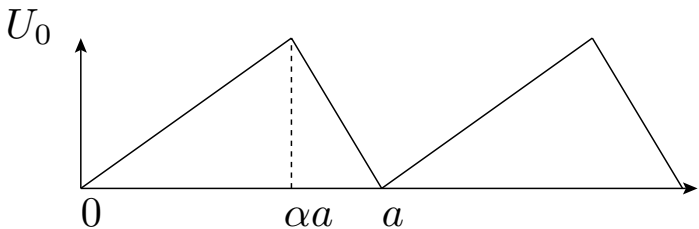}
\caption{Sketch of the sawtooth potential $U(x)$. The potential
has period $a$, $\alpha a$ is
the distance from a minimum to the next maximum on the right, and $U_0$
is the maximum of the potential.} \label{fig_schema}
\end{figure}

We have calculated numerically the function $\Theta(\lambda)$ for the case of the sawtooth potential shown in Fig.~\ref{fig_schema}, with a barrier height $U_0$
of order of several $k_BT$ \cite{dlubensky}.
This function was obtained by first discretizing the operator ${\mathcal L}(\lambda)$ and then calculating its largest eigenvalue using the Ritz variational method.
This method does not require to find a basis specific to the chosen potential, in contrast
to what was done in Ref.~\cite{mehl-seifert08} for the cosine potential. Our numerical method can handle any shape of the potential.

\begin{figure}
\includegraphics[scale=0.4]{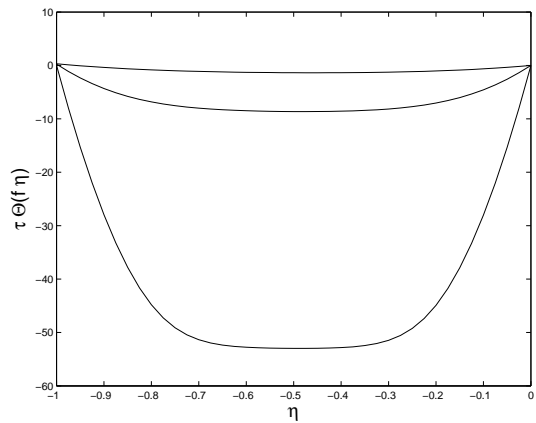}
\caption{Normalized eigenvalue $\tau \Theta(f \eta)$ (with $\tau=a^2/D_0$) as function of $\eta$ for different values of the normalized force $f$; from top to bottom, $f=5$, $f=10$ and $f=20$. The parameters of the potential are $\alpha=0.7$, $U_0/k_B T=5$. The symmetry of all
the curves with respect to $\eta=-1/2$ is Gallavotti-Cohen symmetry expected for model I.
} \label{fig eigenvalue}
\end{figure}

The form of $\Theta(f \eta)$ with $\eta=\lambda/f$ is shown in Fig.~\ref{fig eigenvalue}
for different values of the normalized force $f$.
The symmetry of all the curves with respect to $\eta=1/2$ corresponds
to the symmetry of Eq.~(\ref{GC1}). At weak force, $\Theta(f \eta)$ has a
parabolic shape associated with gaussian fluctuations, whereas at higher
forces a flattening occurs associated with non-gaussian fluctuations \cite{pre,mehl-seifert08,gaspard_andrieux}.
By numerically taking derivatives of $\Theta(\lambda)$
with respect to $\lambda$ near $\lambda=0$, we recover the velocity obtained by
directly solving Eq.~(\ref{eq:FPuncoupled}) \cite{dlubensky,risken}. \\

\begin{figure}
\includegraphics[scale=0.4]{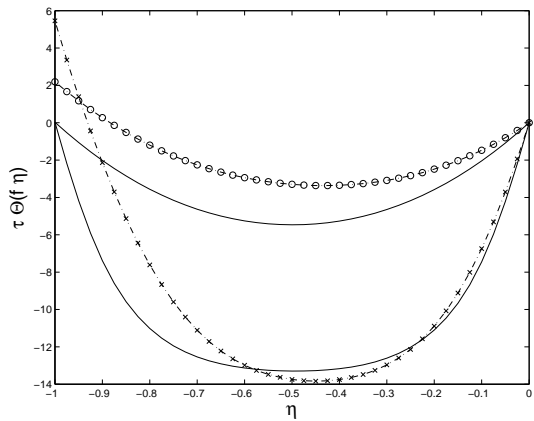}
\caption{Normalized eigenvalue $\tau \Theta(f \eta)$ as function of $\eta$ for a normalized force $f=5$ (top two curves) and $f=10$ (bottom two curves) for the flasing ratchet (model II). The solid curves correspond to the case where the switching rates satisfy detailed balance which leads to the Gallavotti-Cohen symmetry. The curves with circles ($f=5$) and crosses ($f=10$) correspond to cases where detailed balance is broken with constant switching  rates $\omega_1(x)=\omega_2(x)=10 \tau^{-1}$ and with the same potentials. The lack of symmetry in these curves with respect to $\eta=-1/2$ is apparent specially near $\eta=-1$.} \label{fig eigenvalue_switching}
\end{figure}

%
%
We now come to the derivation of the Gallavotti-Cohen symmetry
for the flashing ratchet model (model II).
In this model \cite{armand1,parmeggiani,elston},
the motor has two internal states $i=1,2$, which are described by
two time-independent potentials $U_i(x)$. We assume that these
potentials are periodic with a common period $a$. The probability density
for the motor to be at position
 $x$ at time $t$ and in state $i$ is $P_i(x,t)$. The dynamics of the model is
 described by
 \begin{eqnarray}
 \frac{\partial P_1}{\partial t }  + \frac{\partial J_1}{\partial x}
 &=& - \omega_1(x)  P_1 +   \omega_2(x)  P_2  \nonumber  \\
\frac{\partial P_2}{\partial t }  + \frac{\partial J_2}{\partial
x}
 &=&  \omega_1(x)  P_1  -   \omega_2(x)  P_2  \, ,
 \label{eqs:moteur}
 \end{eqnarray}
where $\omega_1(x)$ and $\omega_2(x)$ are space dependent
transition rates, and the local currents $J_i$ are defined by:
\begin{equation}
      J_i = -D_0 \left(  \frac{\partial P_i}{\partial x}
         +
   \frac{1}{k_B T} \left( \frac{\partial U_i}{\partial x} - F \right) P_i \right),
\end{equation}
with $D_0$ the diffusion coefficient of the motor
and $F$ a non-conservative force acting on the motor.
The transition rates can be modeled using standard kinetics for
the different chemical pathways between the two states of the
motor \cite{parmeggiani}:
 \begin{eqnarray}
        \omega_1(x) &=& [ \omega(x) + \psi(x) e^{\Delta\mu} ] e^{(U_1(x)-fx)/k_B T}, \nonumber \\
        \omega_2(x) &=& [ \omega(x) + \psi(x) ] e^{(U_2(x)-fx)/k_B T},
\end{eqnarray}
where $\Delta \mu=\Delta \tilde{\mu}/k_B T$ is the normalized chemical potential
and $\Delta \tilde{\mu}$ the chemical potential associated with ATP hydrolysis.
Terms proportional to $\omega(x)$ are associated with thermal
transitions, while terms proportional to $\psi(x)$ correspond to transitions induced by ATP hydrolysis. One could easily introduce more chemical pathways than the ones considered here \cite{parmeggiani}
but this extension is not
essential for the present argument. Note that the way the force enters the rates is unambiguous in such a continuous model  \cite{pre,kolomeisky_revue}.

Note that Eq.~(\ref{eqs:moteur})
 can be rewritten as a matrix ${\mathcal  L}$
    of operators:
\begin{eqnarray}
 \frac{\partial }{\partial t }
  \left( \begin{array}{c}
    P_1 \\  P_2
   \end{array}  \right) =  {\mathcal  L}
    \left( \begin{array}{c}
    P_1 \\  P_2
   \end{array}  \right)
 =   \left( \begin{array}{cc}
      {\mathcal L_1} -  \omega_1 &   \omega_2 \\
                   \omega_1 &    {\mathcal L_2} -  \omega_2
   \end{array}  \right)
\left( \begin{array}{c}
    P_1 \\  P_2
   \end{array}  \right) \,\,\,
\label{def:grandL}
\end{eqnarray}
where the action of the  operator  ${\mathcal L_i}$  on a function
$\Phi(x,t)$ is given by
\begin{eqnarray}
  {\mathcal L_i}\Phi = D_0      \frac{\partial^2 \Phi}{\partial x^2} +
   D_0  \frac{\partial }{\partial x} \left(
     \frac{U_i' - F }{k_B T} \Phi    \right).
     \label{def L_i}
 \end{eqnarray}
When $F =0$ and $\Delta\mu =0$, the system is at equilibrium and
\begin{equation}
    \frac{ \omega_2(x)} {\omega_1(x)} =
  \exp\left( \frac{U_2 - U_1}{k_B T}\right) \, .
  \label{usual DB}
\end{equation}
In this case, the stationary solution of the
system~(\ref{eqs:moteur}) is the
 Boltzmann distribution for $P_1$ and $P_2$,
 the currents $J_1$ and $J_2$ vanish and
 there is no global displacement of the motor.
 If both $F$ and  $\Delta\mu$ do not vanish, then the system
 is out of equilibrium and non-vanishing currents  can appear.

%
%
If the switching between the two potentials occurs only
 by thermal transitions, {\it i.e} when $\Delta\mu =0$, the rates
 satisfy the detailed balance condition of Eq.~(\ref{usual DB}),
 even in the presence of a non-zero force $F$.
The Gallavotti-Cohen symmetry follows
by considering a 2x2 diagonal matrix of operators ${\mathcal L}_i(\lambda)$
of the form (\ref{def:Ldeformed}). The symmetry is indeed present as shown in
the solid curves of Fig.~\ref{fig eigenvalue_switching}.
In the general case however, where the normalized force
  $f$ and chemical potential $\Delta \mu$ are both non-zero,
 the relation~(\ref{usual DB}) is no more satisfied and
 the  Gallavotti-Cohen relation (\ref{GC1}) is not valid.
This is shown in the curves with symbols in Fig.~\ref{fig eigenvalue_switching}
where for simplicity we took constant switching rates  $\omega_1=\omega_2=10 \tau^{-1}$. For all the curves of this figure, we took a sawtooth potential $U_1$ with the
same parameters as in Fig.~\ref{fig eigenvalue}, and a potential $U_2$ constant in space. The breaking of the symmetry of Eq.~\ref{GC1} can be interpreted as a result of
the existence of internal degrees of freedom, similarly to the violations
discussed in Ref.~\cite{jarzynski_coarse_graining}.

To establish a fluctuation relation for the flashing ratchet model,
one must consider both the mechanical and chemical currents present \cite{prl,schmiedl-2007}.
\begin{figure}
\includegraphics[scale=0.4]{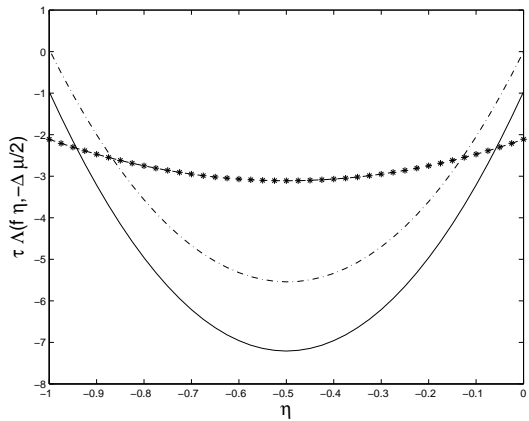}
\caption{For model II, the normalized eigenvalue $\tau \Lambda(f \eta,-\Delta \mu/2)$ is shown as function of $\eta$. The dashed curve corresponds to $f=5$ and $\Delta \mu=0$,  the solid curve corresponds to $f=5$ and $\Delta \mu=10$, the symbols correspond to $f=2$ and $\Delta \mu=10$. The symmetry is recovered in all cases in this description which includes both the mechanical and chemical degrees of freedom.} \label{fig eigenvalue 2 variables}
\end{figure}
Let us introduce the probability density $P_i(x,q;t)$ associated with
 the  probability that at time $t$
 the ratchet is in the internal state $i$, at position $x$ and
 that $q$ chemical units of ATP have been consumed.
 The evolution equations for this probability density is obtained
 by  modifying equations~(\ref{eqs:moteur}) after taking into
 account the dynamics of the discrete variable $q$. We have
 \begin{eqnarray}
 \frac{\partial P_1(x,q,t)}{\partial t} &=&
 \left( {\mathcal L_1}  - \omega_1(x) \right) P_1(x,q,t) \\
&+&  \omega_2^{-1}(x) P_2(x,q+1,t) +  \omega_2^{0}(x) P_2(x,q,t)
  \nonumber  \\
 \frac{\partial P_2(x,q,t)}{\partial t} &=&
 \left( {\mathcal L_2}  - \omega_2(x) \right) P_2(x,q,t) \\
&+&   \omega_1^{0}(x) P_1(x,q,t)
 +
 \omega_1^{1}(x)  P_1(x,q-1,t). \nonumber
 \label{eqs:moteur+chimie}
 \end{eqnarray}
We use a notation similar to that of Ref.~\cite{pre}, where $\omega_i^l(x)$ denotes
the transition rate at position $x$ from
the internal state $i$ with $l=-1,0,1$ ATP molecules consumed. This leads to
$\omega_1^0=\omega e^{(U_1-fx)/k_B T}$, $\omega_2^0=\omega e^{(U_2-fx)/k_B T}$,
$\omega_1^1=\psi e^{(U_1-fx)/k_B T+\Delta \mu}$, $\omega_2^{-1}=\psi e^{(U_2-fx)/k_B T}$,
with $\omega_1(x)=\omega_1^0(x)+\omega_1^1(x)$ and $\omega_2(x)=\omega_2^0(x)+\omega_2^{-1}(x)$.

As above we introduce  two generating functions
   $F_{1,\lambda,\gamma}$ and   $F_{2,\lambda,\gamma}$,
    depending on two parameters $\lambda$ and
 $\gamma$ which are conjugate variables to the position of the ratchet
  and to the ATP counter $q$. We have for $i=1,2$,
  \begin{equation}
   F_{i, \lambda,\gamma}(\zeta, t) =
 \sum_q  \hbox{e}^{\gamma q}  \sum_n  \hbox{e}^{\lambda(\zeta +n)}
 P_i(a(\zeta +n),q; t).
\label{GFMecChim}
 \end{equation}
 The evolution equation for  these generating functions  is obtained
 from Eq.~(\ref{eqs:moteur+chimie}) as
 \begin{equation}
 \frac{\partial }{\partial t}
  \left( \begin{array}{c}
    F_{1,\lambda,\gamma} \\   F_{2,\lambda,\gamma}
   \end{array}  \right)
 =   {\mathcal  L}(\lambda,\gamma)  \left( \begin{array}{c}
    F_{1,\lambda,\gamma} \\   F_{2,\lambda,\gamma}
   \end{array}  \right)  \, ,
\label{EvolMecChim}
 \end{equation}
 with the operator  ${\mathcal  L}(\lambda,\gamma)$  decomposed as
\begin{equation}
 {\mathcal  L}(\lambda,\gamma) = {\mathcal D}(\lambda) + {\mathcal N}(\gamma),
\end{equation}
 with ${\mathcal D}(\lambda)$ the diagonal matrix ${\rm diag}({\mathcal L_1}(\lambda)-\omega_1,{\mathcal L_2}(\lambda)-\omega_2)$, and
 \begin{equation}
{\mathcal  N}(\gamma)  = \left( \begin{array}{cc}
      0 &  \omega_2^0+\omega_2^{-1} e^{-\gamma}  \\
 \omega_1^0+ \omega_1^1 e^{\gamma}
 &   0
   \end{array}  \right).
\label{LdefMecChim}
\end{equation}
Consider now the diagonal matrix $Q$ defined by
${\rm diag} ( e^{-U_1/{k_B T}},e^{-U_2/{k_B T}})$.
By direct calculation, one can check that
$Q^{-1} {\mathcal N}(\gamma) Q = {\mathcal N^\dagger} \left(-\Delta\mu -\gamma \right).$
From Eq.~(\ref{OperatorGC1}), one obtains $Q^{-1} {\mathcal D}(\gamma) Q = {\mathcal D^\dagger}  \left(-\Delta\mu -\gamma \right)$.
By combining these two equations, we conclude that
\begin{equation}
 Q^{-1}  {\mathcal  L}(\lambda,\gamma) Q =
 {\mathcal  L^\dagger}
 \left(-f -\lambda, -\Delta\mu -\gamma \right),
\label{OperatorGCMecaChim}
\end{equation}
which leads to the Gallavotti-Cohen symmetry:
\begin{equation}
 \Lambda(\lambda,\gamma) =
  \Lambda\left(-f -\lambda,  -\Delta\mu -\gamma \right),
\label{FTRatchet2var}
 \end{equation}
 where $\Lambda(\lambda,\gamma)$ is the largest eigenvalue of ${\mathcal  L(\lambda,\gamma)}$.
 If we consider only the mechanical displacement of the ratchet,
 the relevant eigenvalue $\Theta(\lambda)$ is given by
  $\Theta(\lambda)=\Lambda(\lambda,0)$, which clearly does not satisfy the fluctuation relation as shown  in Fig.~\ref{fig eigenvalue_switching}.
In Fig.~\ref{fig eigenvalue 2 variables}, we have computed $\Lambda(f \eta,-\Delta \mu/2)$ for the same potentials and with rates $\omega_i^l(x)$ of the form given above with $\omega(x)=5 \tau^{-1}$ and $\phi(x)=10 \tau^{-1}$.
We have verified that in all cases the symmetry of Eq.~\ref{FTRatchet2var} holds.

In this paper, we have shown that the large deviation function of
the mechano-chemical currents obeys the Gallavotti-Cohen relation.
Another related but different symmetry relation for the entropy
production exists under more general conditions
\cite{kurchan,lebowitz,gaspard_andrieux,maes,mehl-seifert08}. We
have shown here that the symmetry for the currents is valid for
the flashing ratchet model when internal degrees of freedom are
taken into account. This raises a fundamental question concerning
the validity of fluctuations relations and their applicability
 to other types of ratchet models \cite{reiman,hanggi}. Other mechanisms exist which are known to produce
 deviations from fluctuations relations \cite{jarzynski_coarse_graining}, and it would be interesting
 to investigate
whether fluctuations relations can always be restored by a suitable modification of the
dynamics.

On the experimental side, it would be very interesting to
investigate fluctuations relations for molecular motors using
single molecule experiments, in a way similar to what was achieved
in colloidal beads or biopolymers experiments \cite{ritort-2005}.
Using fluorescently labeled ATP molecules, recent experiments with
myosin 5a and with the $F_0-F_1$ rotary motor, aim at simultaneous
recording of the turnover of single fluorescent ATP molecules and
the resulting mechanical steps of the molecular motor \cite{yanagida-2007}. 
These exciting results indicate that a
simultaneous measurement of the values of the mechanical
 and chemical variable of the motor is achievable, and therefore from the
 statistics of such measurements it is possible to construct $P(x,q,t)$.
With enough statistics of such data, one could thus in principle
verify Eq.~\ref{FTRatchet2var}. Such a verification would confirm
that the Gallavotti-Cohen symmetry is a thermodynamic constraint
that plays an essential role in the mechano-chemical coupling of
molecular motors.

We acknowledge fruitful discussions with A. W. C. Lau and J. Prost. D. L.
also acknowledges support from the Indo-French Center CEFIPRA (grant 3504-2).

\bibliographystyle{apsrev}
\bibliography{onsager}

\end{document}